\documentclass[fleqn,twoside]{article}
\usepackage{amsmath}
\usepackage[headings]{espcrc2}
\usepackage{comment}
\usepackage{epsfig}
\usepackage{axodraw}

\readRCS
$Id: espcrc2.tex,v 1.2 2004/02/24 11:22:11 spepping Exp $
\usepackage{graphicx}
\usepackage[figuresright]{rotating}

\newcommand{\plaat}[3]{\raisebox{#2pt}{\mbox{\epsfig{figure=./#1.eps,
width=#3cm}}}}
\newcommand{\plaatw}[3]{\raisebox{#3pt}{\mbox{\epsfig{figure=./#1.eps,
width=#2pt}}}}

\newcommand{\bit}{\begin{itemize}}
\newcommand{\eit}{\end{itemize}}
\newcommand{\nn}{\nonumber}

\newcommand{\bea}{\begin{eqnarray}}
\newcommand{\eea}{\end{eqnarray}}
\def\gb#1{ {\langle #1 ] } }

\title{Recursive equations for arbitrary scattering processes
\thanks{Presented by C.G.Papadopoulos at the 8th DESY Workshop on Elementary Particle Theory,
Loops and Legs in Quantum Field Theory, April 23 -28, 2006,
Eisenach, Germany}}

\author{P.~Draggiotis\address[UoA]{University of Athens,
Physics Department, Nuclear \& Particle Physics Section, \\
$GR-15771$, Athens, Greece}\thanks{pdrag@phys.uoa.gr},
        A.~van Hameren\address[IFJ]{IFJ-PAN Krakow, \\
        31-3420 Krakow, Poland}%
        \thanks{Andre.Hameren@ifj.edu.pl},
        R.~Kleiss\address[IMAPP]{IMAPP, Institute of Mathematics, Astrophysics and Particle Physics,
Radboud University, \\ Nijmegen, the
Netherlands}\thanks{R.Kleiss@science.ru.nl},
A.~Lazopoulos\addressmark[IMAPP]\thanks{A.Lazopoulos@science.ru.nl},
        C.~G.~Papadopoulos\address[NCSR]{Institute of Nuclear Physics, NCSR-DEMOKRITOS
         \\ 15310, Athens,
         Greece}\thanks{costas.papadopoulos@cern.ch},
         M.~Worek\address{Institute of Nuclear Physics, Polish Academy of Sciences
         \\ 31-3420 Krakow, Poland} \thanks{worek@mail.desy.de}
        }

\runtitle{} \runauthor{}

\begin{document}

\begin{abstract}
The usefulness of recursive equations to compute scattering matrix
elements for arbitrary processes is discussed. Explicit results at
tree and one-loop order, obtained by the {\tt HELAC/PHEGAS} package
that is based on the Dyson-Schwinger recursive equations approach,
are briefly presented. \vspace{1pc}
\end{abstract}

\maketitle

\setcounter{footnote}{0}
\section{Introduction}

Recursive equations to compute scattering matrix elements have been
used extensively over the last years in order to obtain results for
multi-leg amplitudes. Their history started essentially with the
work of Berends and Giele~\cite{Berends:1987me}, who were able to
prove the conjectured simple all-$n$ form of Parke and
Taylor~\cite{Parke:1986gb} for the MHV amplitudes in QCD. The
recognition of their usefulness has been expanded recently by the
discovery of a new class of recursive equations, by Britto, Cachazo
and Feng~\cite{Britto:2004ap} and Witten~\cite{Britto:2005fq}.

In this paper we are considering the Dyson-Schwinger (DS) recursive
approach~\cite{Draggiotis:1998gr,Kanaki:2000ey,Kanaki:2000ms,Papadopoulos:2005ky},
and show how this can be used as a general framework for scattering
elements computation. We also present selected results for processes
at tree order and at the one-loop level, obtained with the {\tt
HELAC/PHEGAS}~\cite{Papadopoulos:2000tt} package, which is an
implementation of the DS method.

\section{The Dyson-Schwinger approach}

The traditional representation of the scattering amplitude in terms
of Feynman graphs results to a computational cost that grows like
the number of those graphs, therefore as $n!$ (at tree order), where
$n$ is the number of particles involved in the scattering process.

An alternative\footnote{For other alternatives see
~\cite{Caravaglios:1995cd,Caravaglios:1998yr}.} to the Feynman graph
representation is provided by the Dyson-Schwinger
approach~\cite{Kanaki:2000ms}. Dyson-Schwinger equations express
recursively the $n$-point Green's functions in terms of the
$1-,2-,\ldots,(n-1)$-point functions. In the framework of a theory
with three- and four-point vertices the DS equations are rather
simple and their diagrammatic representation is given below, for
$1\to
n$~\cite{Argyres:1992js,Argyres:1992cp,Argyres:1992kt,Argyres:1992un,Argyres:1993wz,Argyres:2001sa}
amplitude:

\begin{eqnarray}
 && \plaatw{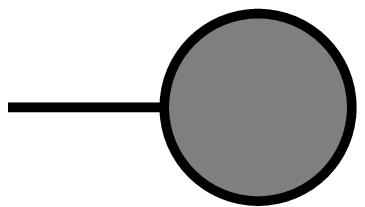}{33}{-5.5} \; =\;\plaatw{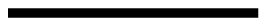}{20}{2}
                       \;+\;           \plaatw{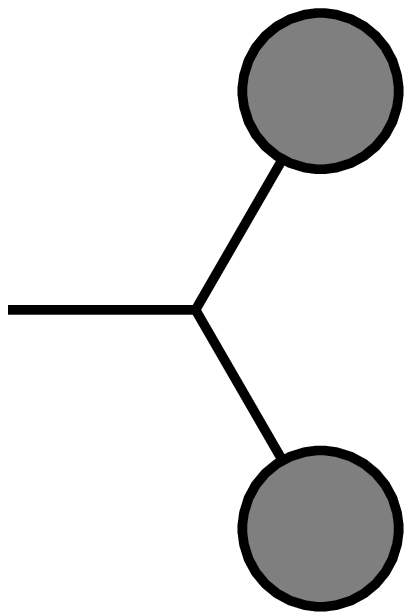}{40}{-28}
                       \;+\; \plaatw{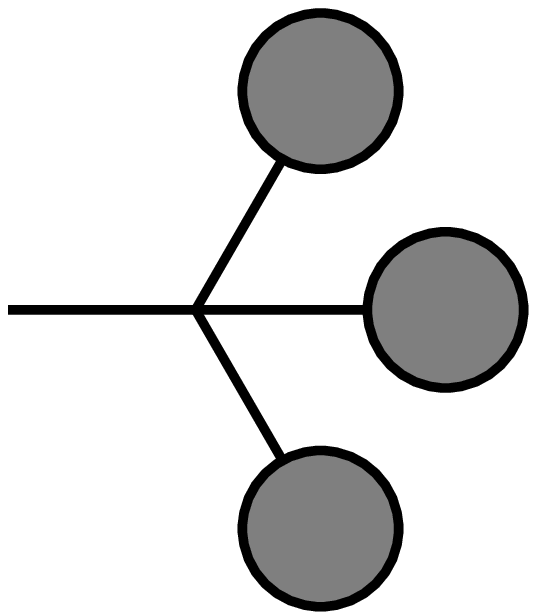}{42}{-22}
\nn \\
                       && \;+\; \plaatw{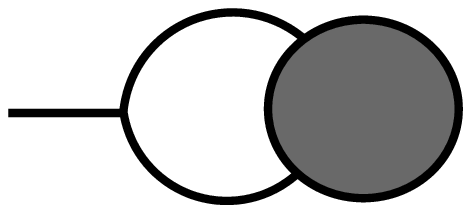}{50}{-7}
                       \;+\; \plaatw{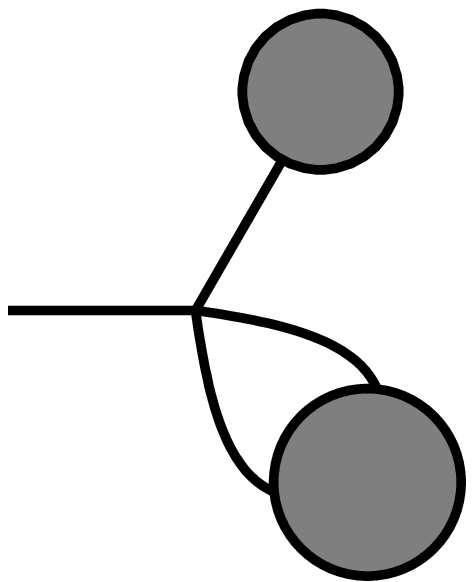}{41}{-20}
                       \;+\; \plaatw{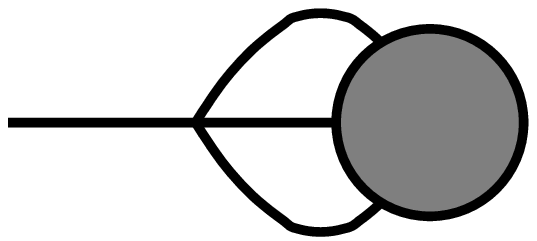}{52}{-8}\;\;
\nn\end{eqnarray}

Omitting the contribution of the second line in the above formula is
equivalent to restrict ourselves at tree order. In order to get an
idea of the actual mathematical form of these equations, let as
consider the simplest case where we are interested to "count
graphs", so by dropping all propagators, couplings, wave-functions,
etc, we end up with the following equation:
\begin{eqnarray}
a(n)= \delta_{n,1} + \frac{1}{2!}\sum \frac{n!}{n_1! n_2!} a(n_1)
a(n_2)
\delta_{n_1+n_2,n} \nn \\
+ \frac{1}{3!} \sum \frac{n!}{n_1! n_2! n_3!} a(n_1) a(n_2) a(n_3)
\delta_{n_1+n_2+n_3,n} \nn\end{eqnarray}

\noindent with the initial condition $a(0)=0$; $a(n)$ is nothing
more than the number of Feynman graphs, contributing to the $1\to n$
matrix element.

The computational cost of DS equations grows like $\sim 3^n$, which
essentially counts the steps used to solve the recursive equations.
Obviously for large $n$ there is a tremendous saving of
computational time, compared to the $n!$ growth of the Feynman graph
approach.

\subsection{Color representation}

Color representation or color decomposition of the amplitude is a
major issue when dealing with multi-parton processes. Let us
consider $n$-gluon scattering with  external momenta
$\{p_{i}\}_1^n$, helicities $\{\varepsilon_{i}\}_1^n$ and colors
$\{a_{i}\}_1^n$  of gluons $i=1,\dots,n$. As is well known the total
amplitude can be expressed as a sum of single trace
terms~\cite{Mangano:1990by}:
\bea
 {\cal
M}(\{p_{i}\}_{1}^{n},\{\varepsilon_{i}\}_{1}^{n},\{a_{i}\}_{1}^{n})
= 2 i g^{n-2} && \nn
\\   \sum^{}_{I\in P(2,\ldots,n)} Tr(t^{a_1}t^{a_{\sigma_I(2)}}\ldots
t^{a_{\sigma_I(n)}}) {\cal A}_I
(\{p_{i}\}_{1}^{n},\{\varepsilon_{i}\}_{1}^{n}) && \nn \eea
where $\sigma_I(2:n)$ represent the $I$-th permutation of the set
$\{2,\ldots,n\}$ and  $Tr(t^{a_1}t^{a_{\sigma_I(2)}}\ldots
t^{a_{\sigma_I(n)}})$ represents a
 trace of generators of the $SU(N_{c})$ gauge group in the fundamental
 representation.  For processes involving quarks a similar but much more cumbersome
 expression can be derived \cite{Mangano:1990by}.

 One of the most interesting aspects of this
 decomposition is the fact that  the ${\cal
 A}_I(\{p_{i}\}_{1}^{n},\{\varepsilon_{i}\}_{1}^{n})$ functions (called dual, partial or
 color-ordered amplitudes), which
 contain all the kinematic information, depend on the permutation and are
 gauge invariant and cyclically
 symmetric in the momenta and helicities of gluons.   The
 color ordered amplitudes are simpler than the full amplitude because
 they only receive contributions from diagrams  with a particular
 cyclic ordering of the external gluons (planar graphs).

Of course to get the full amplitude one has to square the matrix
element,
\bea
\sum_{\{a_i\}_{1}^{n}\{\varepsilon_{i} \}_{1}^{n}} |{\cal
M}(\{p_{i}\}_{1}^{n},\{\varepsilon_{i}
\}_{1}^{n},\{a_{i}\}_{1}^{n})|^2 &&
\nn\\
= g^{2n-4}\sum_{\varepsilon}\sum_{ij}{\cal A}_{I}{\cal C}_{IJ}{\cal
A}_{J}^{*} && \nn
\eea

where the $(n-1)!\times (n-1)!$ dimensional color matrix can be
written in the most general form as follows:
\begin{equation}
{\cal C}_{IJ}= \sum_{1\ldots N_c} Tr(t^{a_1}t^{a_{\sigma_I(2)}}\ldots
t^{a_{\sigma_I(n)}}) Tr(I\leftrightarrow J)^{*}
\end{equation}
There exists a much simpler approach, in fact far superior from the
point of view of an automatized numerical calculation, where the
matrix element is represented as
follows~\cite{'tHooft:1973jz,Kanaki:2000ms,Maltoni:2002mq,Papadopoulos:2005ky},

\bea {\cal
M}(\{p_{i}\}_{1}^{n},\{\varepsilon_{i}\}_{1}^{n},\{c_i,a_i\}_{1}^{n})=
&& \nn
\\ 2 i g^{n-2} \sum^{}_{I=P(2,\ldots,n)} D_I \;\; {\cal A}_I (\{p_{i}\}_{1}^{n},\{\varepsilon_{i}\}_{1}^{n}) &&
\nn \eea

with $c_i,a_i$ the color and anticolor indices for each external
particle, i.e.  $(c,0)$ for quarks, $(0,a)$ for antiquarks, $(c,a)$
for gluons and $(0,0)$ for non-colored particles, and

\[
D_I=\delta_{c_1,a_{\sigma_I(1)}}\delta_{c_2,a_{\sigma_I(2)}}
\ldots\delta_{c_n,a_{\sigma_I(n)}}
\]
or in a more abstract notation
\[
D_I=\delta_{1,\sigma_I(1)}\delta_{2,\sigma_I(2)}
\ldots\delta_{n,\sigma_I(n)}
\]
where $\sigma_I(1:n)$ represent the $I$-th permutation of the set
$\{1,2,\ldots,n\}$. The sequence of numbers $i,\sigma_I(i)$,
$i=1\ldots n$, is identified as a color-connection configuration,
describing the way the color connection is structured. In that
sense, no explicit reference to 'real' color indices is made.
Finally the color matrix takes a very simple form,
\begin{equation}
{\cal C}_{IJ} = N_c^{m(\sigma_I,\sigma_J)}
\end{equation}
where $1\le m(\sigma_I,\sigma_J)\le n$ counts how many common cycles
the permutations $\sigma_I$ and $\sigma_J$ have. For a detailed
description, see \cite{Papadopoulos:2005ky}.

Recursive equations can be written both for the full amplitude,
${\cal M}$, and for the color ordered, ${\cal A}$. In the latter
case the DS equations are identical to the Berends-Giele ones.

For numerical applications the computation of the color ordered
amplitudes suffers from the $n!$ growth related to the number of
color-flow or color-connection configurations. In such cases it is
preferable to write down DS equations for the full amplitude ${\cal
M}(\{p_{i}\}_{1}^{n},\{\varepsilon_{i}\}_{1}^{n},\{c_i,a_i\}_{1}^{n})$
and then perform the incoherent sum
\[ \sum_{c_i, a_i =1\ldots 3 }|{\cal
M}(\{p_{i}\}_{1}^{n},\{\varepsilon_{i}\}_{1}^{n},\{c_i,a_i\}_{1}^{n})|^2
\]
by Monte-Carlo methods. We have recently extended {\tt HELAC} so
that a Monte-Carlo over 'real' colors, or color-configurations can
be performed~\cite{Papadopoulos:2005ky}. A color configuration is
identified by the sequence of numbers $\{c_i,a_i\}_{1}^{n}$,
$c_i,a_i=1\ldots 3$. The details are given in
\cite{Papadopoulos:2005ky}.

Besides the problem related to the color treatment, the summation
over different flavors is also a very important problem when the
flavor of partons at the final state is unidentified, as usually. In
that case a Monte Carlo treatment over flavor degrees of freedom has
been proposed some time ago\cite{Draggiotis:2002hm}, showing that
the  purely gluonic contribution falls from 45.7\% for 3-jet, to
26.6\% for 8-jet production~\cite{Draggiotis:2002hm}.

\subsection{On-shell recursive equations}

During the last year much progress has been made in the
understanding of analytical calculations of color amplitudes in
perturbative Yang-Mills theories. Led by an observation of Witten
\cite{Witten}, Britto, Cachazo and Feng (BCF) have proposed a new
recursion relation for tree amplitudes of gluons
\cite{Britto:2004ap} that naturally arrives at the simplest known
expressions for those amplitudes in terms of Weyl - Van der Waerden
spinors, with Maximal Helicity Violating vertices as building
blocks. Explicit calculations have been performed using this
technique \cite{qcd_calculations}\cite{fermions}, extensions to
amplitudes involving particles from the electroweak sector
\cite{ew_sector} have been pursued and a new approach to one loop
amplitudes has been proposed \cite{one_loop} employing MHV vertices
and unitarity arguments as well as the use of recursive
equations~\cite{Bern:2005cq,Forde:2005hh}.

The BCF recursion relation features some remarkable characteristics,
among which the on-shell analytic continuation of selected off-shell
propagators, the analytic continuation of two selected external
momenta in the complex plane and a decomposition of a color helicity
amplitude into smaller helicity amplitudes with complex external
momenta that doesn't appear to be in direct connection with the
decomposition in Feynman diagrams.

For $n-$gluon amplitude the BCF equation, in a diagrammatic
representation reads as,
\bea {\cal A}(1\ldots n)=
\nn \\
\sum_{j=2}^{n-2}\;\;\;\;\;\frac{\plaat{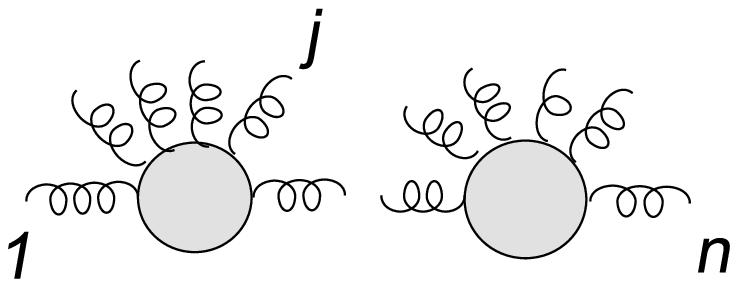}{-15}{4.5}}{P_{1...j}^2}
\nn \eea

or in mathematical terms,

\bea   A_n(1,2,\ldots , (n-1)^-,n^+) = && \cr
\sum_{i=1}^{n-3}\sum_{h=+,-} \left( A_{i+2}({\hat n},1,2,\ldots
i,-{\hat P}^h_{n,i} ) \right.  {1\over P^2_{n,i}} &&
\cr \left. A_{n-i}(+{\hat P}^{-h}_{n,i}, i+1,\ldots , n-2,
{\hat{n-1} } ) \right) && \nonumber\eea
where
\bea  P_{n,i} & = p_n+p_1+\ldots + p_i, \cr \hat P_{n,i} & = P_{n,i}
+{P_{n,i}^2\over \gb{n-1|P_{n,i}|n}} \lambda_{n-1}
\tilde\lambda_{n}, \cr \hat p_{n-1} & =  p_{n-1} -{P_{n,i}^2\over
\gb{n-1|P_{n,i}|n}} \lambda_{n-1} \tilde\lambda_{n} , \cr \hat p_{n}
& =  p_{n} +{P_{n,i}^2\over \gb{n-1|P_{n,i}|n}} \lambda_{n-1}
\tilde\lambda_{n}. \nonumber\eea
where $\lambda_i$ and $\tilde\lambda_i$ are spinors and anti-spinors
coresponding to the momentum $p_i$ and $\gb{i|P|j}\equiv\lambda_i^a
P_{a\dot{b}}\tilde\lambda_j^{\dot{b}}$

The kinematical operation on the momenta of the $1$st and $n$th
particles, is called the 'hat' operation.

It can be proven that BCF equation can be obtained from the
Berends-Giele (or DS) recursive equations, by making use of the
following points~\cite{Draggiotis:2005wq}:

\bit
\item a special gauge choice, that allows the cancelation of all contributions
where diagrams with the first and the last leg meeting in a
three-vertex, as well as with the first and the last leg meeting in
a four-vertex with another external leg. These diagrams are
obviously not-present in the BCF decomposition of the amplitude,
\item a set of relations guaranteed by the kinematical
transformation (the hat 'operation') applied to the chosen momenta,
that exactly takes care of the apparent over-counting of certain
Feynman diagrams within the BCF decomposition,
\item and finally a gauge identity, that relates a 'hatted'
contribution arising from a three vertex, with the un-hatted three-
and four-vertex contributions.
\begin{equation}
\begin{picture}(30,30) (20,-4)
    \SetScale{0.3}
    \SetWidth{0.5}
    \SetColor{Black}
    \Line(15,-4)(90,-4)
    \Line(90,-4)(90,56)
    \Line(90,-4)(150,-4)
    \Line(150,-4)(165,-4)
    \GOval(90,56)(15,15)(0){0.411}
    \COval(90,-4)(2.83,2.83)(45.0){Black}{White}
\Line(88.59,-5.41)(91.41,-2.59)\Line(88.59,-2.59)(91.41,-5.41)
    \COval(90,-4)(2.83,2.83)(45.0){Black}{White}
\Line(88.59,-5.41)(91.41,-2.59)\Line(88.59,-2.59)(91.41,-5.41)
    \COval(89,-3)(11.31,11.31)(-45.0){Black}{White}
\Line(94.66,-8.66)(83.34,2.66)\Line(83.34,-8.66)(94.66,2.66)
  \end{picture}
  =
\begin{picture}(32,30) (5,-4)
    \SetScale{0.3}
    \SetWidth{0.5}
    \SetColor{Black}
    \Line(15,-4)(90,-4)
    \Line(90,-4)(90,56)
    \Line(90,-4)(150,-4)
    \Line(150,-4)(165,-4)
    \GOval(90,56)(15,15)(0){0.411}
  \end{picture}
\;\;\; \;\;\; +\;\;\;
 \begin{picture}(30,30) (5,-4)
    \SetScale{0.3}
    \SetWidth{0.5}
    \SetColor{Black}
    \Line(15,-4)(90,-4)
    \Line(90,-4)(90,56)
    \Line(90,-4)(150,-4)
    \Line(150,-4)(165,-4)
    \GOval(90,56)(15,15)(0){0.411}
    \Line(90,-4)(90,-45)
    \Text(30,-15)[lb]{\small{$\epsilon_{\sigma}$}}
    \COval(90,-4)(8.83,8.83)(60.0){Black}{White}
  \end{picture}
\nonumber
\end{equation}
where
\begin{equation}
\begin{picture}(45,30) (10,-5)
    \SetScale{0.3}
    \SetWidth{0.5}
    \SetColor{Black}
    \Line(15,-4)(90,-4)
    \Line(90,-4)(90,56)
    \Line(90,-4)(150,-4)
    \Line(150,-4)(165,-4)
    \GOval(90,56)(15,15)(0){0.411}
    \Line(90,-4)(90,-45)
    \Text(30,-15)[lb]{\small{$\epsilon_{\sigma}$}}
    \COval(90,-4)(8.83,8.83)(60.0){Black}{White}
  \end{picture}=-z V_{\mu\nu\rho\sigma}
\nonumber \end{equation}
with $V_{\mu\nu\rho\sigma}$ the QCD four-vertex.

\eit

Although BCF equations are very powerful in order to obtain
'analytical' results, their numerical implementation does not show
up any real gain as compared to the Berends-Giele (or DS) ones.  In
fact for a moderate number of external particles $8<n<12$ their
complexity and therefore their CPU-time consumption is substantially
larger than that of the Berends-Giele equations\footnote{See
also~\cite{Dinsdale:2006sq}.}. Nevertheless their usefulness in
computing tree- as well as one-loop amplitudes is still an
unexplored territory.

\section{{\tt HELAC/PHEGAS}: results}

{\tt HELAC/PHEGAS} is a computer package that incorporates the DS
approach to compute scattering cross section for arbitrary process.
In the computation of the matrix elements, it includes all Standard
Model particles and interactions, both in Feynman and Unitary
gauges. There are two options to deal with the colored particles,
namely the color-connection approach, in which all color ordered
amplitudes are computed, and the color-configuration approach in
which the full amplitude is given, followed by a Monte-Carlo
treatment of the color summation. In the phase space generation and
integration, sector, {\tt PHEGAS} is using a multi-channel approach,
each Feynman graph, identified with a potential generation channel,
followed by an optimization of the a priori weights, entering the
calculation of the global phase-space density. Moreover for
multi-particle processes where the number of Feynman graphs makes
the use of a multi-channel approach impossible, other phase-space
generation methods and packages, like {\tt
HAAG}~\cite{vanHameren:2002tc} and {\tt
DURHAM}~\cite{Papadopoulos:2005ky} are used.

{\tt HELAC/PHEGAS} has been used extensively to produce physically
relevant results\footnote{See for
instance~\cite{Berends:2000gj,Gleisberg:2003bi}.}. In the sequel we
are going to restrict ourselves to two specific examples, in order
to reveal its potential for physics studies.

The first example refers to the process $p\;p \to t \;\bar{t} \;b\;
\bar{b} \;b \;\bar{b} $. From the physics point of view it consists
the irreducible  background of $t\bar{t} H H $ production, which
seems interesting in a high-luminosity LHC (SLHC) version, for
studying HHH coupling~\cite{Gianotti:2002xx}. From the computational
point of view, it is a challenging process, and a nice example to
demonstrate the ability of {\tt PHEGAS/HELAC} to deal with QCD
processes in a realistic setup.

The number of Feynman graphs contributing to this process is {\bf
1454} (for a $gg$ initial state), with 5! color-connection
configurations. We have used the structure functions and $\alpha_s$
from {\tt PDFLIB}, CTEQ-4L (LO). Kinematical decays of $t \to b W^+$
have been implemented and the following set of cuts has been used:
$p^{b}_{T}>20 GeV$, $ |\eta_b| < 2.5$, $\Delta R > 0.5$. The result
for the total cross section is 1.053 $\pm$ 0.073 (fb) @ LHC energy.

The second example is the computation of Fermion-Loop (FL)
contributions in six-fermion production processes in $e^+e^-$
collisions~\cite{Argyres:1995ym,Beenakker:1996kn,Beenakker:2003va}.
It is an explicit example of the use of DS equations to compute
one-loop amplitudes. This is achieved in a rather straightforward
way, by adding to the tree-order SM vertices, contributions arising
from 1PI graphs at one loop~\cite{Hollik:2004dz,vanHameren}. In such
a way using {\tt FORM} we were able to calculate all $V_1V_2V_3$ and
$V_1V_2V_3V_4$ vertices at one loop, for arbitrary kinematical
configuration (all particles off-shell) and for all gauge bosons
$V=\gamma, Z, W^\pm$. Then, as is dictated by the well known
quantum-field theoretic argument, the amplitude is just the
tree-order one with the tree-order vertices replaced by the
generalized one-loop vertices.

We take as an example the process $e^- e^+ \to \mu^- \bar{\nu}_\mu u
\bar{d} \,\tau^- \tau^+ $. The number of Feynman Graphs contributing
to the process is 208, whereas the number of DS vertices (the steps
needed to compute the amplitude) is 140. We have used the following
set of kinematical cuts: $ E_l,E_q
>5$GeV and $ m_{ll}, m_{qq} > 10$GeV, and the result for $E=500$GeV
is given by $\sigma_0/ab=54,96(26)$ at tree order
$\sigma_1/ab=57,31(28)$ at the one loop, with a $K$-factor given by
$K/100=4.28(2)$. The MC data for this particular run are as follows:
MC points generated: 1 Million (961792), MC points used after cuts:
404842, real time of running:6 1/2 hours on a very basic PC.

\section{Outlook}

Recursive equations have been proven to be the framework for an
efficient matrix element computation for arbitrary scattering
processes. They are the basic ingredients towards the construction
of an automatized generator including NLO corrections. The fusion
with parton-shower generators and the understanding of the working
of this fusion, will be one of the main tasks in the near future.
Precision calculations will offer the solid basis needed for
discoveries in future high-energy colliders.

\section*{Acknowledgments}
C.G.P acknowledges support by the EU Transfer of Knowledge programme
MTKD-CT-2004-014319. A. van H. and M.W. acknowledge support from the
Greek-Polish bilateral programme of GSRT for 2004-2006. P.D. is
currently co-funded by the European Social Fund (75\%) and National
Resources (25\%)-EPEAEK B! - PYTHAGORAS.

\end{document}